\documentclass[twocolumn,fleqn,10pt]{wlscirep}

\usepackage[utf8]{inputenc}
\usepackage[T1]{fontenc}
\usepackage{multirow, booktabs, tabularx, longtable}
\usepackage{multirow, makecell}

\usepackage{tabulary,longtable,afterpage}
\usepackage{booktabs}
\makeatletter
\newcommand{\aftertwo}[1]{\afterpage{\if@firstcolumn #1
  \else\afterpage{#1}\fi}}
\makeatother

\title{Local impacts on road networks and access to critical locations during extreme floods}

\author[1,2,*]{Simone Loreti}
\author[3]{Enrico Ser-Giacomi}
\author[1,2]{Andreas Zischg}
\author[1,2,4,5]{Margreth Keiler}
\author[6,7,*]{Marc Barthelemy} 
\affil[1]{University of Bern, Institute of Geography, Bern, 3012, Switzerland}
\affil[2]{University of Bern, Oeschger Centre for Climate
Change Research, Mobiliar Lab for Natural Risks, Bern, 3012, Switzerland}
\affil[3]{Massachusetts Institute of Technology, Department of Earth, Atmospheric and Planetary Sciences, Cambridge MA, 02139, United States}
\affil[4]{University of Innsbruck, Department of Geography, Innsbruck, 6020, Austria}
\affil[5]{Austrian Academy of Sciences, Institute of Interdisciplinary Mountain Research, Innsbruck, 6020, Austria}
\affil[6]{Institut de Physique Th\'eorique, CEA,
CNRS-URA 2306, Gif-surYvette, F-91191, France}
\affil[7]{Centre d’Analyse et de Math\'ematique Sociales (CNRS/EHESS), Paris, 75006, France}

\affil[*]{To whom correspondence should be addressed. E-mail: simone.loreti@giub.unibe.ch or marc.barthelemy@ipht.fr}

\begin{abstract}

\vspace{-0.5cm}
Floods affected more than 2 billion people worldwide from 1998 to 2017 and their occurrence is expected to increase due to climate warming, population growth and rapid urbanization. Recent approaches for understanding the resilience of transportation networks when facing floods mostly use the framework of percolation but we show here on a realistic high-resolution flood simulation that it is inadequate. Indeed, the giant connected component is not relevant and instead, we propose to partition the road network in terms of accessibility of local towns and define new measures that characterize the impact of the flooding event. Our analysis allows to identify cities that will be pivotal during the flooding by providing to a large number of individuals critical services such as hospitalization services, food supply, etc.  This approach is particularly relevant for practical risk management and will help decision makers for allocating resources in space and time.

\end{abstract}

\begin{document}

\flushbottom
\maketitle
\thispagestyle{empty}

\vspace{-0.5cm}
\noindent 
\textbf{Date:}
31 January 2022.\\
\textbf{Disclaimer:}
This research article has been published the 28th of January 2022 on \href{https://www.nature.com/articles/s41598-022-04927-3 }{Scientific Reports}.\\
\textbf{DOI:} https://doi.org/10.1038/s41598-022-04927-3.\\
\textbf{Cite this article:}
Download citation from the \href{https://www.nature.com/articles/s41598-022-04927-3#citeas}{Scientific Reports} citation tool.

\section*{Introduction}
\label{Introduction}

Floods are the most frequent and life-threatening natural hazards related disasters \cite{AghaKouchak,Doocy,Kreibich,Wallemacq} and are expected to increase in occurences and damages due to a large variety of factors such as climate warming with intensification of precipitation extremes, population growth, and rapid urbanization \cite{Alfieri,Paprotny,SalmanAbdullahi,Tanoue,Ukkusuri2021,Visser,Wasko,Winsemius}. 
\textcolor{black}{As other natural disasters, floods reduce accessibility and serviceability of the road transportation network, which is one of our most}
\textcolor{black}{valuable infrastructure assests}
\textcolor{black}{\cite{Berdica2002}.}
\textcolor{black}{These}
\textcolor{black}{disruptive effects on the road system are intimately related to the vulnerability and resilience of the transportation network and are traditionally analysed with a number of methodologies, such as systems dynamics models, stochastic and optimization processes \cite{Goncalves2020}, network science \cite{Goncalves2020, Pan2021, Gajanayake2020,Mattsson2015}, demand and supply models \cite{Mattsson2015, Gajanayake2020}, and approaches based on traffic data \cite{Pan2021}.}
\\
\\
\textcolor{black}{
Network science is a multidisciplinary field \cite{Latora2017, Molontay2021, Newman2018,Vespignani2018} which offers robust and elegant tools and measures to investigate the topological features and properties of a network \cite{Albert2001,Newman2003}. In network science, a}
well-established framework for assessing the robustness of any kind of disrupted network is percolation theory \cite{Boccaletti2014,Gao2012b,Newman2018}, which was recently applied to road networks, to investigate the disruptive effects of
\textcolor{black}{
earthquakes \cite{Aydin2018a,Aydin2018b,Dong2019,Zhou2019b} and traffic congestion \cite{Hamedmoghadam2021,Li2015,Zeng2019,Zeng2020}, as well as of flooding events \cite{Abdulla2019_conference_paper,Abdulla2020_conference_paper,Abdulla2020_SIS,Abdulla2021,Fan2020,Farahmand2020,Dong2020a,Dong2020b,Ganin,Wang2019,Yadav}.
Abdulla et al. \cite{Abdulla2019_conference_paper} modelled the flood propagation over one of the neighborhoods road network of Houston, U.S, as a percolation mechanism. In the proposed process, they randomly removed (flooded) the network's nodes based on their elevation and their vicinity to already flooded nodes}
\textcolor{black}{(using Baye's rule).}
\textcolor{black}{Despite}
\textcolor{black}{they mentioned to simulate}
\textcolor{black}{
a percolation process, which would traditionally include the computation of the giant connected component, they actually calculated the Latora \& Marchiori's global efficiency \cite{Latora} as a measure of network connectivity.
Both Abdulla \& Birgisson \cite{Abdulla2020_conference_paper} and Abdulla et al. \cite{Abdulla2020_SIS} investigated the potential failure of the Houston road network undergoing fluvial floods, by employing different removal sequences of nodes in a flood-related percolation process. For the entire set of nodes they first calculated four centrality measures, i.e. degree, betweenness, closeness and eigenvector centrality, which then served to rank their nodes.} 
\textcolor{black}{Following}
\textcolor{black}{the nodes ranking from the highest to the lowest rank, the authors performed the nodes' removal sequence for each centrality measure and measured the network connectivity through the giant component size. Eventually, they found that the greatest damage to the network was caused by the removal sequence associated to the betweenness centrality measure,}
\textcolor{black}{
in agreement with results about the resilience of air-transportation weighted networks (see for example Dall'Asta et al \cite{DallAsta2006}).}
\textcolor{black}{
Abdulla et al. \cite{Abdulla2020_SIS} also presented a flood diffusion model and investigated the effects of the initial conditions of the flood diffusion on the road network connectivity. To mimic different initial conditions of the flood, the authors performed the flood diffusion from different sets of nodes, correspondent to those nodes which ranked the highest values of centrality measures, i.e. degree, betweenness, closeness and eigenvector centrality. By measuring the size of the giant component for each fraction of removed (flooded) nodes, they found that the road networks connectivity was particularly vulnerable to the flood propagation which started from nodes with high values of betweenness. 
\textcolor{black}{
However, their framework did not include any hydrological or hydraulic model, nor was validated.}
Abdulla \& Birgisson \cite{Abdulla2021} examined the impact of random and targeted disruptions on the Houston road network. To represent the perturbative action of floods on the road network they used the same mechanism of targeted disruption as in Abdulla et al. \cite{Abdulla2019_conference_paper}. Through the calculation of the giant component size, 
\textcolor{black}{
they compared the disruptive effects of targeted and random failure on the road network. For low percentage of nodes removal (i.e. between 15\% and 30\%), they found that the network was more robust against targeted failure than towards random one, while for a higher percentage of nodes removal (i.e. between 30\% and 70\%), the network was instead more robust against random failure than targeted disruptions.}
The also found similar connectivity profiles (giant components size) by varying the road network size. They 
\textcolor{black}{concluded by emphasizing}
the suitability of the giant component in assessing the robustness of road networks.
Fan et al. \cite{Fan2020} presented a percolation-based diffusion model of flood propagation and recession, similar to the susceptible-exposed-infected-recovered (SEIR) disease transmission model. However, their model was not built up directly over the road network, but over an overlapping \quotes{network grid}, composed of squared cells and each of them containing a different number of road segments. Therefore, the flood propagation 
\textcolor{black}{occurred}
among grid cells, with a probability of flooding proportional to the number of neighbouring flooded cells and to a constant transmission rate. With a similar idea, they also simulated the recession phase of the flood. 
Despite the mention to a percolation-based process, the authors did not calculated the giant connected component or the equivalent size of outbreak \cite{Newman2018}, but they characterised the entire process of flood diffusion mainly by calculating the fraction of flooded cells. In addition, all the model parameters, i.e. the transmission rate, the recovery rate, and the exposed rate were estimated through a curve fitting with real data.
Farahmand et al. \cite{Farahmand2020} proposed a probabilistic approach to the failure of road networks (inversely proportional to the distance between roads and channels network, and directly proportional to a channel's vulnerability function), which was assessed through the calculation of the giant component size of the network. During the process of nodes removal, they found dramatic decreases (leaps) in the connectivity profile represented by the giant component size, suggesting the presence of critical roads.}
Dong et al. \cite{Dong2019} proposed a probabilistic links removal to mimic earthquakes-induced failures and addressed the corresponding effects on the Portland (USA) road network through a non-conventional percolation process. Instead of using the traditional giant connected component, they introduced \quotes{robust components} defined as the union of connected components in a network, containing at least one critical facility (as a hospital), showing a two-phase transition. In their perspective, the giant connected component doesn't fairly represent the functionality of a network undergoing such a disaster, claiming that measures of network’s robustness should consider critical infrastructures.
In another work, 
\textcolor{black}{
Dong et al. \cite{Dong2020a} employed the same idea of robust component for assessing the robustness of a road network affected by a flood, resulting in an almost gradual transition.}
\textcolor{black}{
They simulated the fluvial flood in the entire county by assigning a probability of failure to each link, which was proportional to the distance between the link and the floodway (but not considering any flood depth). For each value of link removal probability, they calculated the robust component and observed a complete shutter of the robust component at a threshold of $p_c=0.8$. They also reported a sudden drop of the robust component when $2\%$ of the roads were inundated ($p=0.02$), resulting in a loss of accessibility to hospitals for a $22\%$ of the network.}
Even if 
\textcolor{black}{the}
papers 
\textcolor{black}{related to the \quotes{robust components}}
might apparently represent a step towards non-percolation approaches, these authors are still removing gradually an increasing fraction of nodes from zero to one and monitor a cluster size for each fraction of removed nodes. Instead of monitoring the size of the largest connected cluster as in the traditional percolation, they monitor the size of the disjoint union of all clusters containing at least a hospital and consider it as if it were a single large component. The hospitals serve uniquely in this work as markers for constructing the clusters that will constitute the robust component.
\textcolor{black}{
Dong et al. \cite{Dong2020b} proposed to apply a standard percolation process, i.e. with a random removal of the network nodes, for assessing the robustness of twelve U.S. cities and three states's road networks, when subject to disaster-induced failures. 
For all the cities and states networks, they showed a discrepancy between the results generated with a simulation-based approach and an analytical one, i.e. the generating function formalism \cite{Callaway2000, Newman2001}, due to the spatially embedded nature of the road networks (indeed, by decreasing the node degree assortativity, the discrepancy diminished). 
Therefore, they claimed about the inadequacy of the generating functions formalism for assessing the robustness of road networks. 
However, the Dong et al. \cite{Dong2020b} study and conclusions were limited to the random failure mechanism of the network, which is far from real disruptive processes observed in some natural disasters, as floods.}
\textcolor{black}{
Inspired by percolation theory, Ganin et al. \cite{Ganin} analysed the increase of traffic delay in six U.S. cities's road networks with the increased severity of the road links disruptions, by employing a mechanism of random links removal. A low, medium and high percentage of disrupted road links represented, respectively, accidents (low), power failures or sever flooding events (medium), and snow, ice, or dust storms (high). The authors, therefore, used the traffic delay for assessing the towns resilience, as alternative quantity to the traditional giant component size.}
Wang et al. \cite{Wang2019} compared three types of disruptions to both USA and China’s road networks: random, flood-induced and localized failures. They presented the flood-induced failure as a new typology of network disruption, with intermediary features between a random and a localized disruption\textcolor{black}{,} and resulting in an abrupt first-order phase transition. 
\textcolor{black}{
Yadav et al. \cite{Yadav} studied different failure scenarios on the London Rail Networks (LRN) system, both as a supra-single network and a multi-layer interdependent network \cite{Kivela2014, Bianconi2018}. The failure mechanisms included a random failure of the overall network, a local random failure, a targeted failure based on centrality measures and one scenario representing a flood event. The flood-like failure consisted in dividing the nodes into three groups based on their proximity to the river Thames and, within each group, randomly removing them from the network.}
\\
\\
In \textcolor{black}{most of} these previous studies 
\textcolor{black}{related to the application of percolation-based flood modeling on the road system}
\cite{Abdulla2019_conference_paper,Abdulla2020_conference_paper,Abdulla2020_SIS,Abdulla2021,Fan2020,Farahmand2020,Dong2020a,Dong2020b,Ganin,Wang2019,Yadav}, the network's functionality was expressed through the size of the largest connected component $P_\infty$ and its evolution when roads are removed.
\textcolor{black}{
A summary of those studies and their key features related to the percolation-based process is provided in} 
\textcolor{black}{\cref{tab:table_literature_review}.} 
However, $P_\infty$ is an aggregated quantity that does not capture the entire network's information and more importantly, does not reflect the local reality. Therefore, \textcolor{black}{more precise information is necessary, in particular at a local scale,} for a realistic evaluation of the disruption's effects. 
Here, we present a realistic and extreme flood scenario \cite{Zischg} based on physically plausible rainfall scenarios and a high spatio-temporal resolution, and show that it cannot be described adequately by a percolation transition and the behavior of the giant component. 
Indeed, the giant component size $P_{\infty}$ never reaches zero during the flooding process and the percolation threshold is never reached. This demonstrates the need for alternative useful measures of the impact of flooding events on the road network. Here we propose such measures and partition the network into \emph{towns}, through a Voronoi tessellation, which allows us to extend the concept of \emph{network's functionality} to the entire network and at a local level.  This approach led us to define time-dependent measures alternative to percolation, both at a local and at a global scale, which provide a realistic assessment of a flood-induced disruptive event on a transportation network. In order to demonstrate the relevance of these new measures, we use them on a realistic flood simulation and also compare the results to a null random model.

%
%

\aftertwo{
\onecolumn
\begingroup
\setlist[itemize]{label={--},nosep, leftmargin=*,
before=\vspace*{-\baselineskip}}
\setlength{\extrarowheight}{1.5pt}
%
\begin{small}


\begin{longtable}
{m{2.9cm} m{1.3cm} m{2.4cm} m{2.1cm} m{4.5cm} m{1.7cm}}

\toprule
$Reference$
& $Figure(s)$
& $x-axis$
& $y-axis$ 
& $Failure \; mechanism$
& $Place$\\ 
\midrule
\endhead


Abdulla et al. \cite{Abdulla2019_conference_paper} 
& $4$ 
& $0<p<0.5$ 
& efficiency \cite{Latora} 
& 
\begin{itemize}
\item probabilistic, 
based on nodes elevation and vicinity to flooded nodes 
\vspace*{-\baselineskip}
\end{itemize}
& Houston \\
\cline{1-6}


Abdulla \linebreak 
\& Birgisson \cite{Abdulla2020_conference_paper}
& $3$ 
& $0 < p < 1$ 
& $0 < P_\infty < 1$ 
& 
\begin{itemize}
\item random
\item betweenness
\item degree
\item closeness
\item eigenvector
\vspace*{-\baselineskip}
\end{itemize}
& Houston \\ 
\cline{1-6}

\multirow{2}{*}[-3em]{Abdulla et al. \cite{Abdulla2020_SIS}}

& $7$
& $0 < p < 0.2$ 
& 
\begin{itemize}[leftmargin=0pt]
\renewcommand\labelitemi{}
\item $0.6 < P_\infty < 1$
\item $0.5 < P_\infty < 1$
\item $0.6 < P_\infty < 1$
\item $0.3 < P_\infty < 1$
\item $0.45 < P_\infty < 1$
\vspace*{-\baselineskip}
\end{itemize}
& 
\begin{itemize}
\item diffusion (random seed)
\item diffusion (betweenness seed)
\item diffusion (degree seed)
\item diffusion (closeness seed)
\item diffusion (eigenvector seed)
\vspace*{-\baselineskip}
\end{itemize}
& \multirow{2}{*}[-2.4em]{Houston} \\
\cline{2-5}

& $8$ 
& $0 < p < 0.2$
&
\begin{itemize}[leftmargin=0pt]
\renewcommand\labelitemi{}
\item $0.2 < P_\infty < 1$
\item $0.05 < P_\infty < 1$
\item $0.4 < P_\infty < 1$
\item $0.4 < P_\infty < 1$
\item $0.5 < P_\infty < 1$
\vspace*{-\baselineskip}
\end{itemize}
& 
\begin{itemize}
\item random
\item betweenness
\item degree
\item closeness
\item eigenvector
\vspace*{-\baselineskip}
\end{itemize}
&
\\
\cline{1-6}


Abdulla \linebreak 
\& Birgisson \cite{Abdulla2021}
& 4,7,8,9,10 
& $0 < p < 1$ 
& $0 < P_\infty < 1$ 
& 
\begin{itemize}
\item random
\item fluvial flood proxy as in \cite{Abdulla2019_conference_paper} 
\vspace*{-\baselineskip}
\end{itemize}
& Houston \\
\cline{1-6}


Fan et al. \cite{Fan2020}
& 3
& $0 < t < 10$ 
& c(t)
& 
\begin{itemize}
\item probabilistic SEIR model
\vspace*{-\baselineskip}
\end{itemize}
& Harris County, U.S.A. \\
\cline{1-6}


Farahmand et al. \cite{Farahmand2020}
& $4$ 
& $0 < p < 1$ 
& $0 < P_\infty < 1$ 
& 
\begin{itemize}
\item probabilistic, 
based on road-channel proximity and channel's vulnerability
\vspace*{-\baselineskip}
\end{itemize}
& Harris County, U.S.A. \\
\cline{1-6}


Dong et al. \cite{Dong2020a}
& $4$ 
& $0 < p < 1$ 
& $0 < RC < 1$ 
& 
\begin{itemize}[noitemsep,topsep=0pt,leftmargin=*]
\renewcommand\labelitemi{--}
\item probabilistic, 
based on the distance between road and floodway
\vspace*{-\baselineskip}
\end{itemize}
& Harris County, U.S.A. \\
\cline{1-6}


Dong et al. \cite{Dong2020b}
& $8,9,11$ 
& $0 < p < 1$ 
& $0 < P_\infty < 1$ 
& 
\begin{itemize}[noitemsep,topsep=0pt,leftmargin=*]
\renewcommand\labelitemi{--}
\item random (generating functions)
\item random (simulations)
\vspace*{-\baselineskip}
\end{itemize}
& $12$ cities, $3$ states in U.S.A. \\
\cline{1-6}


Ganin et al. \cite{Ganin}
& $5$ 
& $0 < p < 1$ 
& traffic delay 
& 
\begin{itemize}[noitemsep,topsep=0pt,leftmargin=*]
\renewcommand\labelitemi{--}
\item probabilistic, based on the road length
\vspace*{-\baselineskip}
\end{itemize}
& $6$ cities in U.S.A. \\
\cline{1-6}


\multirow{2}{*}[-21.5em]{Wang et al. \cite{Wang2019}}

& $3 \, {(a{-}b)}$
& ${0 < 1{-}p < 0.8}$ 
& 
\begin{itemize}[noitemsep,topsep=0pt,leftmargin=0pt]
\renewcommand\labelitemi{}
\item $0 < P_\infty < 1$
\item $0 < P_\infty < 1$
\item $0 < P_\infty < 1$
\vspace*{-\baselineskip}
\end{itemize}
& 
\begin{itemize}[noitemsep,topsep=0pt,leftmargin=*]
\renewcommand\labelitemi{--}
\item random (from entire network)
\item normal flood
\item localised (random walk)
\vspace*{-\baselineskip}
\end{itemize}
& U.S.A., China \\
\cline{2-6}

& $4 \, {(b)}$
& $5 < R < 285$ 
& 
\begin{itemize}[noitemsep,topsep=0pt,leftmargin=0pt]
\renewcommand\labelitemi{}
\item $0.93 < P_\infty < 1$
\vspace*{-\baselineskip}
\end{itemize}
& 
\begin{itemize}[noitemsep,topsep=0pt,leftmargin=*]
\renewcommand\labelitemi{--}
\item normal flood
\vspace*{-\baselineskip}
\end{itemize}
& New York \\
\cline{2-6}

& $5$ 
& $5 < R < 285$ 
&
\begin{itemize}[noitemsep,topsep=0pt,leftmargin=0pt]
\renewcommand\labelitemi{}
\item $0 < P_\infty < 1$
\item $0 < P_\infty < 1$
\item $0 < P_\infty < 1$
\item $0 < P_\infty < 1$
\item $0 < P_\infty < 1$
\item $0.2 < P_\infty < 1$
\item $0.7 < P_\infty < 1$
\item $0 < P_\infty < 1$
\vspace*{-\baselineskip}
\end{itemize}
& 
\begin{itemize}[noitemsep,topsep=0pt,leftmargin=*]
\renewcommand\labelitemi{--}
\item not specified
\vspace*{-\baselineskip}
\end{itemize}
&
\begin{itemize}[noitemsep,topsep=0pt,leftmargin=0pt]
\renewcommand\labelitemi{}
\item Guangxi
\item Henan
\item Sichuan
\item Hunan
\item Florida
\item Illinois
\item Michigan
\item Minnesota
\vspace*{-\baselineskip}
\end{itemize} \\
\cline{2-6}

& Sup. $2$ 
& ${0 < 1{-}p < 0.8}$ 
&
\begin{itemize}[noitemsep,topsep=0pt,leftmargin=0pt]
\renewcommand\labelitemi{}
\item $0 < P_\infty < 1$
\item not specified
\item $0 < P_\infty < 1$
\item $0 < P_\infty < 1$
\item $0.1 < P_\infty < 1$
\vspace*{-\baselineskip}
\end{itemize}
& 
\begin{itemize}[noitemsep,topsep=0pt,leftmargin=*]
\renewcommand\labelitemi{--}
\item random (from entire network)
\item normal floods
\item uniform random floods
\item pearson-III random floods
\item localised (random walk)
\vspace*{-\baselineskip}
\end{itemize}
& 
U.S.A., China \\
\cline{2-6}

& Sup. $3$ \textbf{*}
&  
\begin{itemize}[noitemsep,topsep=0pt,leftmargin=0pt]
\renewcommand\labelitemi{}
\item ${0 < 1{-}p < 0.04}$
\item ${0 < 1{-}p < 0.06}$
\item ${0 < 1{-}p < 0.06}$
\item ${0 < 1{-}p < 0.15}$
\item ${0 < 1{-}p < 0.08}$
\vspace*{-\baselineskip}
\end{itemize}
&
\begin{itemize}[noitemsep,topsep=0pt,leftmargin=0pt]
\renewcommand\labelitemi{}
\item $0 < P_\infty < 1$
\item not specified
\item $0.75 < P_\infty < 1$
\item $0.75 < P_\infty < 1$
\item $0.8 < P_\infty < 1$
\vspace*{-\baselineskip}
\end{itemize}
& 
\begin{itemize}[noitemsep,topsep=0pt,leftmargin=*]
\renewcommand\labelitemi{--}
\item random (from entire network)
\item normal floods
\item uniform random floods
\item pearson-III random floods
\item localised (random walk)
\vspace*{-\baselineskip}
\end{itemize}
&
\begin{itemize}[noitemsep,topsep=0pt,leftmargin=0pt]
\renewcommand\labelitemi{}
\item Guangxi
\item Henan
\item Hunan
\item Sichuan
\item Zhejiang
\vspace*{-\baselineskip}
\end{itemize} \\
\cline{2-6}

& Sup. $4$ \textbf{*}
&  
\begin{itemize}[noitemsep,topsep=0pt,leftmargin=0pt]
\renewcommand\labelitemi{}
\item ${0 < 1{-}p < 0.03}$
\item ${0 < 1{-}p < 0.04}$
\item ${0 < 1{-}p < 0.04}$
\item ${0 < 1{-}p < 0.03}$
\item ${0 < 1{-}p < 0.03}$
\item ${0 < 1{-}p < 0.04}$
\item ${0 < 1{-}p < 0.05}$
\item ${0 < 1{-}p < 0.03}$
\item ${0 < 1{-}p < 0.08}$
\vspace*{-\baselineskip}
\end{itemize}
&
\begin{itemize}[noitemsep,topsep=0pt,leftmargin=0pt]
\renewcommand\labelitemi{}
\item $0 < P_\infty < 1$
\item not specified
\item $0.9 < P_\infty < 1$
\item $0.9 < P_\infty < 1$
\item $0.6 < P_\infty < 1$
\vspace*{-\baselineskip}
\end{itemize}
& 
\begin{itemize}[noitemsep,topsep=0pt,leftmargin=*]
\renewcommand\labelitemi{--}
\item random (from entire network)
\item normal floods
\item uniform random floods
\item pearson-III random floods
\item localised (random walk)
\vspace*{-\baselineskip}
\end{itemize}
&
\begin{itemize}[noitemsep,topsep=0pt,leftmargin=0pt]
\renewcommand\labelitemi{}
\item Florida
\item Illinois
\item Iowa
\item Michigan
\item Minnesota
\item New York
\item Ohio
\item Tennessee
\item Texas
\vspace*{-\baselineskip}
\end{itemize} \\
\cline{2-6}

& Sup. $7$
& $5 < R < 285$ 
& 
\begin{itemize}[noitemsep,topsep=0pt,leftmargin=0pt]
\renewcommand\labelitemi{}
\item $0.75 < P_\infty < 1$
\vspace*{-\baselineskip}
\end{itemize}
& 
\begin{itemize}[noitemsep,topsep=0pt,leftmargin=*]
\renewcommand\labelitemi{--}
\item normal flood
\vspace*{-\baselineskip}
\end{itemize}
& Sichuan \\
\cline{2-6}

& Sup. $9$
& $5 < R < 285$ 
&
\begin{itemize}[noitemsep,topsep=0pt,leftmargin=0pt]
\renewcommand\labelitemi{}
\item $0.2 < P_\infty < 1$
\item $0.2 < P_\infty < 1$
\item $0.2 < P_\infty < 1$
\item $0.4 < P_\infty < 1$
\item $0.1 < P_\infty < 1$
\item $0.1 < P_\infty < 1$
\vspace*{-\baselineskip}
\end{itemize}
& 
\begin{itemize}[noitemsep,topsep=0pt,leftmargin=*]
\renewcommand\labelitemi{--}
\item not specified
\vspace*{-\baselineskip}
\end{itemize}
&
\begin{itemize}[noitemsep,topsep=0pt,leftmargin=0pt]
\renewcommand\labelitemi{}
\item Zhejiang
\item Florida
\item Minnesota
\item New York
\item Tennessee
\item Texas
\vspace*{-\baselineskip}
\end{itemize} \\
\cline{1-6}


\multirow{2}{*}[-0.8em]{Yadav et al. \cite{Yadav}}

& 
\multirow{4}{*}[1em]{$2$}
& 
\multirow{4}{*}[1em]{${0 < p < 1}$ }
& 
\begin{itemize}[noitemsep,topsep=0pt,leftmargin=0pt]
\renewcommand\labelitemi{}
\item $0 < P_\infty < 1$
\item $0 < P_\infty < 1$
\item $0 < P_\infty < 1$
\vspace*{-\baselineskip}
\end{itemize}
& 
\begin{itemize}[noitemsep,topsep=0pt,leftmargin=*]
\renewcommand\labelitemi{--}
\item random
\item degree
\item betweenness
\vspace*{-\baselineskip}
\end{itemize}
& 
\begin{itemize}[noitemsep,topsep=0pt,leftmargin=0pt]
\renewcommand\labelitemi{}
\item LR (sing.)
\item SF (single)
\item ER (single)
\vspace*{-\baselineskip}
\end{itemize}
\\
\cline{4-6}

&
& 
& 
\begin{itemize}[noitemsep,topsep=0pt,leftmargin=0pt]
\renewcommand\labelitemi{}
\item $0 < P_\infty < 1$
\item $0.1 < P_\infty < 1$
\item $0.1 < P_\infty < 1$
\vspace*{-\baselineskip}
\end{itemize}
& 
\begin{itemize}[noitemsep,topsep=0pt,leftmargin=*]
\renewcommand\labelitemi{--}
\item random
\vspace*{-\baselineskip}
\end{itemize}
& 
\begin{itemize}[noitemsep,topsep=0pt,leftmargin=0pt]
\renewcommand\labelitemi{}
\item LR (multi)
\item SF (multi)
\item ER (multi)
\vspace*{-\baselineskip}
\end{itemize}
\\
\cline{4-6}

&
& 
& 
\begin{itemize}[noitemsep,topsep=0pt,leftmargin=0pt]
\renewcommand\labelitemi{}
\item $0 < P_\infty < 1$
\item $0.1 < P_\infty < 1$
\item $0.1 < P_\infty < 1$
\vspace*{-\baselineskip}
\end{itemize}
& 
\begin{itemize}[noitemsep,topsep=0pt,leftmargin=*]
\renewcommand\labelitemi{--}
\item degree
\vspace*{-\baselineskip}
\end{itemize}
& 
\begin{itemize}[noitemsep,topsep=0pt,leftmargin=0pt]
\renewcommand\labelitemi{}
\item LR (multi)
\item SF (multi)
\item ER (multi)
\vspace*{-\baselineskip}
\end{itemize}
\\
\cline{4-6}

&
& 
& 
\begin{itemize}[noitemsep,topsep=0pt,leftmargin=0pt]
\renewcommand\labelitemi{}
\item $0 < P_\infty < 1$
\item $0.1 < P_\infty < 1$
\item $0.1 < P_\infty < 1$
\vspace*{-\baselineskip}
\end{itemize}
& 
\begin{itemize}[noitemsep,topsep=0pt,leftmargin=*]
\renewcommand\labelitemi{--}
\item betweenness
\vspace*{-\baselineskip}
\end{itemize}
& 
\begin{itemize}[noitemsep,topsep=0pt,leftmargin=0pt]
\renewcommand\labelitemi{}
\item LR (multi)
\item SF (multi)
\item ER (multi)
\vspace*{-\baselineskip}
\end{itemize}
\\
\cline{2-6}

& 
$3$
&  
\multirow{2}{*}[1em]{\shortstack[l]
{
${0 < n_{rm} < 65}$
\\
\\
\\
(removal from\\flood risk maps)
}}
& 
\begin{itemize}[noitemsep,topsep=0pt,leftmargin=0pt]
\renewcommand\labelitemi{}
\item $0.7 < P_\infty < 1$
\item $0.5 < P_\infty < 1$
\item $0.7 < P_\infty < 1$
\item $0.1 < P_\infty < 1$
\item $0.1 < P_\infty < 1$
\vspace*{-\baselineskip}
\end{itemize}
& 
\begin{itemize}[noitemsep,topsep=0pt,leftmargin=*]
\renewcommand\labelitemi{--}
\item flood (river proximity + rand.)
\item random-global
\item random-local
\item targeted after flood
\item targeted before flood
\vspace*{-\baselineskip}
\end{itemize}
& LR \\
\cline{2-6}

& 
\multirow{2}{*}[-1.5em]{$5$}
& 
\multirow{2}{*}[-1em]{\shortstack[l]
{
${0 < n_{rm} < 31}$
\\
\\
\\
(removal from\\flood risk maps)
}}
& 
\begin{itemize}[noitemsep,topsep=0pt,leftmargin=0pt]
\renewcommand\labelitemi{}
\item $0.8 < P_\infty < 1$
\item $0.7 < P_\infty < 1$
\item $0.4 < P_\infty < 1$
\vspace*{-\baselineskip}
\end{itemize}
& 
\begin{itemize}[noitemsep,topsep=0pt,leftmargin=*]
\renewcommand\labelitemi{--}
\item flood (river proximity + random)
\vspace*{-\baselineskip}
\end{itemize}
& 
\begin{itemize}[noitemsep,topsep=0pt,leftmargin=*]
\renewcommand\labelitemi{--}
\item undergr.
\item overgr.
\item DLR
\vspace*{-\baselineskip}
\end{itemize}
\\
\cline{4-6}

& 
& 
& 
\begin{itemize}[noitemsep,topsep=0pt,leftmargin=0pt]
\renewcommand\labelitemi{}
\item $0.2 < P_\infty < 1$
\item $0.6 < P_\infty < 1$
\item $0 < P_\infty < 1$
\vspace*{-\baselineskip}
\end{itemize}
& 
\begin{itemize}[noitemsep,topsep=0pt,leftmargin=*]
\renewcommand\labelitemi{--}
\item targeted after flood
\vspace*{-\baselineskip}
\end{itemize}
& 
\begin{itemize}[noitemsep,topsep=0pt,leftmargin=*]
\renewcommand\labelitemi{--}
\item undergr.
\item overgr.
\item DLR
\vspace*{-\baselineskip}
\end{itemize}
\\

\bottomrule
\caption{
\textcolor{black}{This is a summary of the reviewed articles which discuss the application of percolation processes to road networks, as proxies of flooding events. In the table header, \emph{Figure(s)} indicates the Figure's number in the reference paper which is related to the percolation process, while \emph{x-axis} and \emph{y-axis} represent the axes of the percolation process's graphic. The x-axis covers a number of roles, as fraction of removed nodes, $p$, time, $t$ (days), surface runoff, $R$ ($\frac{mm}{day}$), and number of removed nodes, $n_{rm}$, while the y-axis mostly represents the giant connected component, $P_{\infty}$.
In a few cases the percolation-based process (y-axis) is represented by the global efficiency of Latora \& Marchiori \cite{Latora}, or by the Fan et al. \cite{Fan2020} fraction of flooded grid cells, $c(t)$, by the Dong et al. \cite{Dong2019} robust component, \emph{RC}, or by the traffic delay in Ganin et al. \cite{Ganin}. About the last two columns, i.e the \emph{failure mechanism} and \emph{place}, please refer to the corresponding reference papers for further details.
Note to the reader (\textbf{*}): both in Figure Sup. $3$ and in Sup. $4$, in Wang et al. \cite{Wang2019}, each range of $1{-}p$ values refers to the chinese province or american state on the same table row (Guangxi, Henan, Hunan, etc., or Florida, Illinois, Iowa, etc.), while each range of $P_{\infty}$ values refers to the failure mechanism on the same table row. However, to relate the unique range of $P_{\infty}$, here shown in this table, to all the chinese provinces mentioned in Sup. $3$ (if reading Figure Sup. $3$), or to all the american states mentioned in Sup. $4$ (if reading Figure Sup. $4$), the minimum value of $P_{\infty}$ was selected among those ones associated to the provinces or states in \cite{Wang2019} and shown here as the overall lower endpoint of $P_{\infty}$. Therefore, for each failure mechanism, the corresponding range of $P_{\infty}$ contains the minimum value of $P_{\infty}$ among all the values of the chinese provinces or american states.}
}
\label{tab:table_literature_review} 
\\
\end{longtable}
%
\end{small}
\endgroup
\twocolumn
}

\section*{Results}
\label{Results}

\subsection*{Flooding is not a percolation transition}
\label{Temporal_percolation}

\begin{figure*}
\centering
\includegraphics[width=11.4cm]{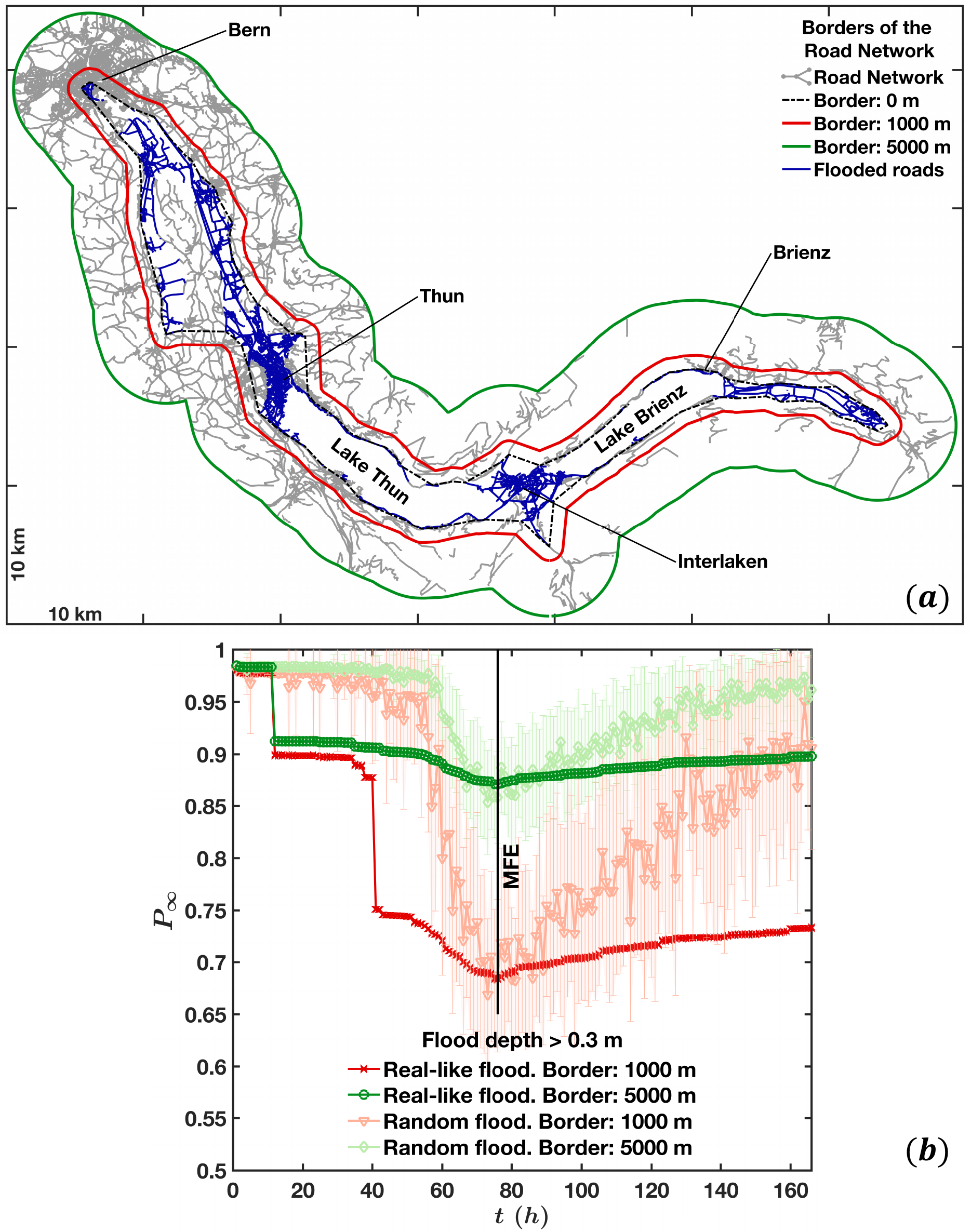}
\caption{\textbf{(a)} Definition of borders. Given the set of endpoints (nodes) of the flooded roads at the maximum flood extension, a tight $2-$D boundary \cite{MATLABboundary} surrounding those points was drawn. The boundary was created with a shrink factor \cite{MATLABboundary} equal to $s_f = 0.5$, which is between a convex hull ($s_f = 0$) and the \quotes{tightest single-region boundary} ($s_f = 1$). That boundary corresponds to the dash-dot line called \quotes{border: $0$ meters}. Larger boundaries, corresponding to \quotes{border: $1000$ meters} and \quotes{border: $5000$ meters}, were created by radially expanding \cite{MATLABpolybuffer} the \quotes{border: $0$ meters} by a distance of $1000$ and $5000$ meters, respectively.
\textbf{(b)} Evolution during the flood of the fraction  $P_\infty$ of nodes belonging to the giant connected component computed for both real-like floods \citep{Zischg} and randomly generated floods and for both the $1000$ and $5000$ meter-borders shown in \textbf{(a)}. The curves representing the random floods were computed by averaging the values of $20$ realizations. Here, they are shown with $1-$standard deviation error bars.
\textcolor{black}{More precisely, those average values resulted from the arithmetic mean performed over the $20$ values of $P_\infty$ (related to the $20$ realizations of random floods), at each time-step, while the vertical bars around the average values represented the standard deviation performed over the $20$ values of $P_\infty$, still at each time step. Therefore, both mean and standard deviation (1-std) of the $20$ values of $P_\infty$ are here shown, for each time step (obviously, those $20$ values are different at each time step). This figure was produced with Matlab \cite{MATLAB2019}.} 
}
\label{fig:borders}
\end{figure*}

In order to test our approach and our new metrics, we will consider a portion of the Swiss road network built up from the \emph{swissTLM3D} dataset 
\cite{Swisstopo} and shown in \cref{fig:borders}(a). This area comprises the major towns of Bern, Thun, Interlaken and Brienz and two Swiss lakes, the Thun Lake and the Brienz Lake. This portion of road network was selected \textcolor{black}{in such a way to include} the largest extension of the disrupted zone, represented by the \textcolor{black}{blue} flooded roads, and a surrounding area (limited by the \textcolor{black}{green} perimeter) where the effects of the disruptions were expected to occur as well. We use a realistic flood model \cite{Zischg} based on physically plausible rainfall scenarios\textcolor{black}{,} with a high spatio-temporal resolution (a spatial resolution of $50$ m and temporal resolution of a hour) \textcolor{black}{and we model the} impact of the rainfall in a deterministic way\textcolor{black}{,} based on physical laws \textcolor{black}{and} ground morphology and properties. In addition, this model takes into account both flood protection's works and the elevation difference between the river's water surface and the bridge over it. In contrast \textcolor{black}{to} other simplified models such as in
\textcolor{black}{
\cite{Abdulla2019_conference_paper,Abdulla2020_conference_paper,Abdulla2020_SIS,Abdulla2021,Fan2020,Farahmand2020,Dong2020a,Dong2020b,Ganin,Wang2019,Yadav}}
, the model here takes into account all \textcolor{black}{the} relevant field features
\textcolor{black}{
(additional information in the \cref{flood_failure} $-$ \nameref{flood_failure}).}
The flooded roads (in blue in \cref{fig:borders}(a)) result from these simulations \cite{Zischg}, while both the green and red external borders were created from an alpha-shape perimeter of the flooded roads (with shrinking factor $0.5$) by broadening the perimeter outwards over a distance of $5000$ m and $1000$ m, respectively (see \cref{fig:borders}(a)).
We compare our results to a random null model where we remove the roads at random\textcolor{black}{, i.e.} by randomly reshuffling the water levels \textcolor{black}{simulated by Zischg et al. \cite{Zischg} over the entire set of roads}. In this way, the null model has the same number of flooded roads per hour as in the real-like floods, allowing the comparison between the (time-dependent) random-like \textcolor{black}{flood} and the real-like \textcolor{black}{one}. In the following, we will coin the simulation result as a \emph{real-like flood} and the null model as a \emph{random flood}.

For these flooding processes, we first monitor the quantity $P_\infty$ (see \cref{fig:borders}(b))\textcolor{black}{,} defined as the fraction of nodes that belongs to the giant connected component. The simulations start at time $t_0=0$ hours where there is no flood, the network is intact and $P_\infty$ is maximum and equal to $1$. During the flooding, an increasing number of roads is covered by water until the maximum flood extension \textcolor{black}{(MFE) is} reached at $t_{MFE}=75$ hours. After this peak, the flood withdraws and \textcolor{black}{the} flooded roads eventually dry up, becoming accessible again\textcolor{black}{. We} monitor this \quotes{recovery} phase until $t=166 \;h\approx$ 1 week. Note that not all flooded roads were removed from the network, but only those \textcolor{black}{ones} covered by more than $0.3$ meters\textcolor{black}{,} which corresponds to the height at which vehicles start to float \cite{Pregnolato}. The results presented in \cref{fig:borders}(b) first show that \textcolor{black}{the magnitude of the $P_\infty$'s variations is affected by the choice of the border delimiting the flooded area},
which points to the problem of defining the area of study\textcolor{black}{. Nevertheless, the $P_\infty$'s} minimum is reached at the same time $t_{MFE}$ for all choices of borders and for both \textcolor{black}{flood scenarios}. \textcolor{black}{A recovery phase then follows, characterised by a slow growth of $P_\infty$, which yet doesn't reach the initial stage of $P_\infty=1$}. The curves corresponding to the random flood grow faster that those \textcolor{black}{ones} obtained for the real-like flood, mostly due to the retention effect of lakes which slows down the flood \textcolor{black}{withdrawal}.
\textcolor{black}{
Indeed, the flood duration for the roads adjacent to the lakes is higher than for those ones which are only affected by river flooding. Although we added additional time to the simulations to consider the slow process of lake emptying, the full recovery would need more time.}
\textcolor{black}{Focusing on the real-like case, we observe jumps of $P_\infty$ during the flooding process ($t<t_{MFE}$),} which are due to the presence of natural elements\textcolor{black}{,} such as lakes or mountains, that constrain the road network\textcolor{black}{. Flooding of those constrained roads has then} a dramatic impact on the giant component size and creates these discontinuities in $P_\infty$. The existence of these discontinuities depends obviously on the border definition: for the $5000 \; m$ case, the behavior of $P_\infty$ is much smoother as there are multiple paths connecting nodes in this larger network. Again, this rises the question \textcolor{black}{on} the choice of the embedding region when studying a percolation related quantity such as $P_\infty$. Also, we observe that these discontinuities are washed out in the random case\textcolor{black}{,} leading to an artificially smooth evolution of the giant component size. We don't see jumps in the recovery phase \textcolor{black}{since} those roads whose disruption would cause the jumps in the \textcolor{black}{flood's} growing phase have not been dried up yet (or completely) at the last time step of the simulations. \textcolor{black}{In previous studies where floods were considered in the framework of percolation,}
\textcolor{black}{as in 
\cite{Abdulla2021,Dong2020a,Farahmand2020,Wang2019},
we note that the authors reported a complete (or nearly complete) disintegration of the giant connected component, showing the decrease of $P_\infty$ (or $RC$ in \cite{Dong2020a}) from one to zero. This process was carried out by removing a fraction of nodes from zero to one in \cite{Abdulla2021,Dong2020a,Farahmand2020} and from zero to around $0.8$ in \cite{Wang2019}.}
This is in sharp contrast with our observations for both real-like and random floods\textcolor{black}{, where} $P_\infty$ undergoes a reduction of less than $30\%$ (depending on the choice of borders)\textcolor{black}{. This would suggest} that percolation might not be the correct framework\textcolor{black}{,} and $P_\infty$ not the most relevant quantity\textcolor{black}{,} for understanding the impact of flooding on transportation networks.

\begin{figure*}
\centering
\includegraphics[width=11.4cm]{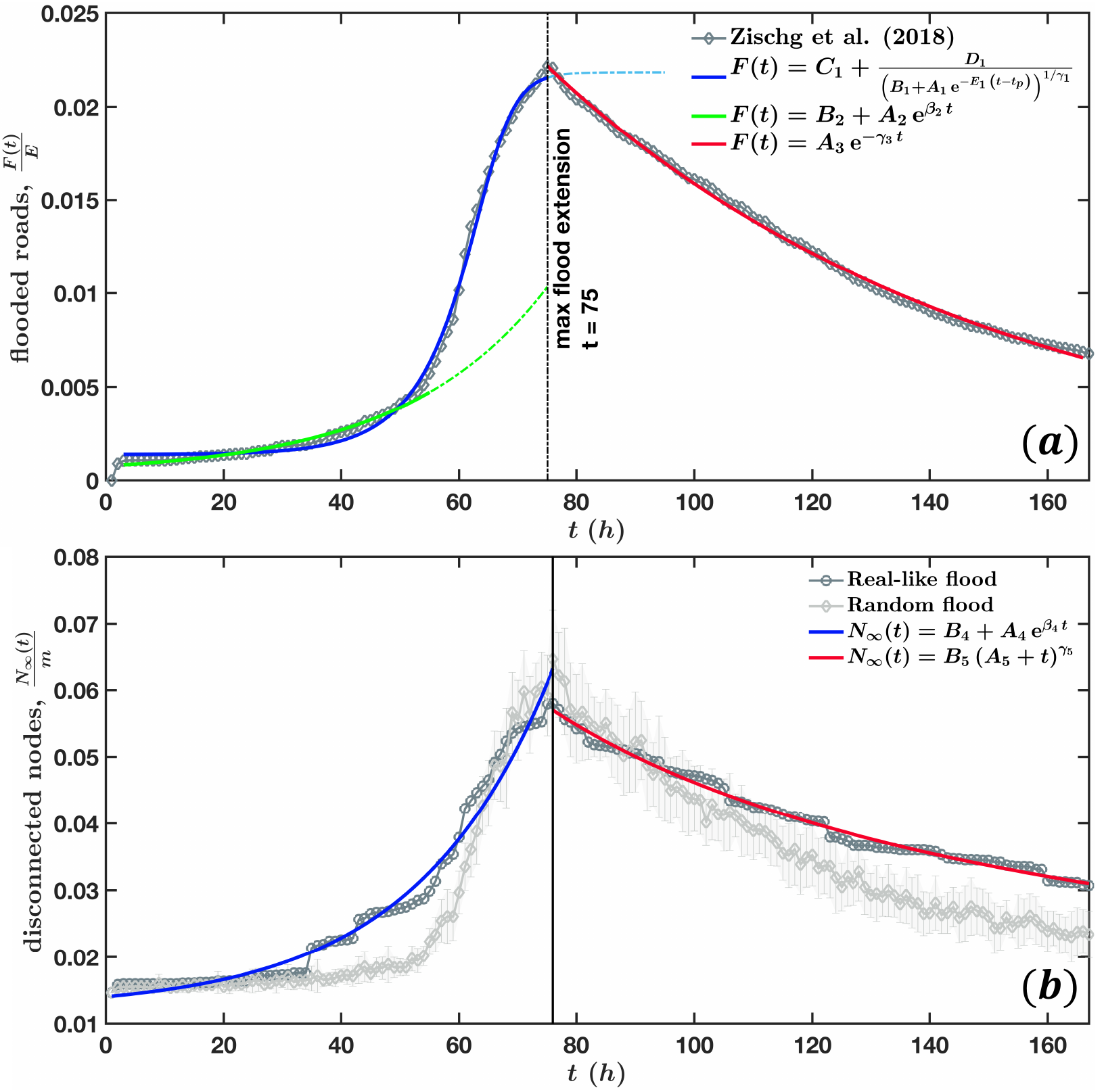}
\caption{\textbf{(a)} Time evolution of the number of flooded roads, $F(t)$. During the Zischg et al. \citep{Zischg} simulations, roads were marked as flooded if the water depth was greater than $0.3$ meters, which is considered as the limit of water's depth for driving a car (with average ground clearance) \cite{Pregnolato}. In terms of number of flooded roads, the earliest times of the flooding event are well characterised by an exponential growth. However, this function does not describe the saturation phase here observed, which is instead typical of logistic-like functions. A fair approximation of the entire growing phase is given by a generalised logistic function, which fits particularly well the saturation phase, up to the maximum flood extension. The recovery phase, that is the phase of flood  withdraw, follows instead a slow exponential decay. \textbf{(b)} Time evolution of the number of disconnected nodes, $N_\infty(t)$. This metric is here illustrated for both the real-like flood simulated by Zischg et al. \cite{Zischg} and the randomly generated flood. For the real-like flood, the dynamics of $N_\infty(t)$ is well approximated by an exponential growth and decay, respectively for the increasing and recovery phases of the flood. Although both $N_\infty(t)$ - calculated for the two types of floods -, reach a similar peak at $t_{MFE}$, the number of nodes which disconnect from the rest of the network is generally larger for real-like floods than in random floods, in both phases of flood's increase and withdraw. This occurs since real-like floods are more spatially concentrated (i.e. less dispersed) than random floods, with a higher probability of surrounding and isolating nodes. Therefore, modelling and simulating a flood event with a random blockage of roads would be misleading, generally resulting in an underestimation of $N_\infty(t)$, in several moments of a flooding event.
\textcolor{black}{This figure was produced with Matlab \cite{MATLAB2019}.}}
\label{fig:number_flooded_roads_linlin_crop}
\end{figure*}

In addition, we note that \textcolor{black}{a flood event} displays different dynamics. Indeed, during the simulation we can also compute for each hour the number of flooded roads, $F(t)$, \textcolor{black}{which} resulting curve is shown in \cref{fig:number_flooded_roads_linlin_crop}(a). The curve displays a peak at the maximum flood extension $t_{MFE}$ and divides the flood into two phases, the growing phase ranging from $t=0$ to $t_{MFE}$ and the recovery phase (or flood's withdrawal phase) for $t>t_{MFE}$. Interestingly enough, we note that the fraction of disconnected nodes is smaller than $7\%$\textcolor{black}{,} which is significantly smaller than what is usually assumed in percolation type approaches.
The beginning of the growing phase can be well fitted by an exponential function with a time scale of order $14$ hours. The end of this phase displays a saturation effect and is better fitted by a generalized logistic function (see
\textcolor{blue}{Supplementary Table 1}
for details on the various fits \textcolor{black}{and 
\textcolor{blue}{Supplementary Text}
for details on the generalised logistic function}). These results suggests that the evolution of the number of flooded roads could be described - at least in the first 
phase - by a generalized Verhulst's equation  \cite{Verhulst} of the form
\begin{equation}
\begin{split}
    \frac{dF(t)}{dt} = \beta_1 \Bigl(F-C_1\Bigr) &\left(1 - \left( \frac{F-C_1}{D_1} \right)^{\gamma_1} \right) \\ 
    &\quad \text{for \; $t_0<t<t_{MFE}$} 
\end{split}
%
\label{eq:generalized_Verhulst_equation}
\end{equation}
where the \textcolor{black}{intrinsic} growth rate is here $\beta_1\approx 0.145$. %
The recovery phase is well fitted by an exponential decay with a time scale of order $75$ hours\textcolor{black}{,} consistent with the observed slow decay. It would be interesting to construct a simple model for deriving these equations and results\textcolor{black}{, to better} understand the critical parameters governing these time scales. 
\textcolor{black}{
Indeed, the form of these equations might suggest a more complex flood propagation model on road networks than those ones presented in \cite{Abdulla2020_SIS,Fan2020}. In addition, the Abdulla et al. \cite{Abdulla2020_SIS} model does not include any hydrological or hydraulic component, nor a validation, while the Fan et al. \cite{Fan2020} model, which is applied to grid cells instead of roads, does not consider the water depth, nor the difference in height among the nodes (in their flooding probability).}
Another related measure is the number of disconnected nodes \textcolor{black}{in the whole network}, $N_\infty(t)$ (defined in \cref{infinitedistances} in the Methods), \textcolor{black}{which} time evolution is shown in \cref{fig:number_flooded_roads_linlin_crop}(b) for both flood scenarios. \textcolor{black}{Despite the amount of flooded roads per hour is the same in both flood scenarios, we observe a faster increase of $N_\infty(t)$ during the growing phase and a slower decrease during the recovery phase in the real-like flood, than in the random flood.}
\textcolor{black}{This difference is due to the retention effect of the lakes, which implies a storage of flood water and a consequent time lag in releasing the same water.} 
\textcolor{black}{Therefore, important roads located along the right and left shorelines of both lakes (and connecting the upstream with downstream areas)} are inundated \textcolor{black}{earlier} in time than the rest of the basin and are drying up much later because the lakes are going back to their normal level very slowly (much more slowly than the river floodplains). These differences highlight once again the danger of using simple random model for describing floods.

\subsection*{Realistic measures of the local impact of flooding}

We saw above that describing the flooding process by percolation could actually be misleading and that the results could depend on a variety of factors such as the choice of borders for example. Maybe more importantly, we note that during floods, individuals will tend to reach the closest town centers  where they can find (at least the basic) supplies, services or assistance. When the closest town cannot be reached anymore due to flooding, individuals will have to reroute to the next closest town in order to fulfill their needs. The rerouting process from a town to another can be viewed as a varying \quotes{attraction basin} of the neighbouring town, as illustrated in \cref{fig:rerouting_mechanism}. Some towns will see their attraction basin decrease, while others will gain some nodes and attract additional individuals. In order to capture these important effects that occur during disruptive events, we first identify the nodes that correspond to towns and identify their attraction basin as their Voronoi cell (for details, see Methods). The Voronoi cell of a town $c$ will be denoted as $K_c(t)$ and is defined as $K_c(t) \doteq \{ i \mid \min\limits_{c'} (d_{i c'} (t)) = d_{ic} (t) \}$ (its size is denoted by $|K_c(t)|$). The shortest distance $d_{ic}(t)$ from a node $i$ to the town $c$ is computed using Dijkstra's 
algorithm over the graph of non-flooded roads at time $t$. We can then define the town's loss $L_c(t)=1-|K_c(t)|/|K_c(t_0)|$ (see Methods for details on these metrics). These metrics are shown in \cref{fig:newmetrics}(a,b,c) for each of the $67$ towns composing the network. 

The evolution of $|K_c(t)|$ over the entire flood period (\cref{fig:newmetrics}(a)) displays both losses of nodes for certain towns and an increase of their attractivity for others. In order to visualize in \cref{fig:newmetrics}(a) which towns 
undergo a size reduction or increase, we represent them in different colors according to their behavior. We find that about half of the towns ($54 \%$) are not affected by the flood, about $31 \%$ undergo an overall size reduction, and about $15 \%$ display an overall size increase. Although we do not have any spatial information from \cref{fig:newmetrics}(a), the distribution of non affected towns indicates that the flood impact on the transportation network is limited within the area defined in \cref{fig:borders} (i.e. within a perimeter of $5000 \; m$), and does not involve the entirety of the network here considered. A second information we can infer from the affected towns is about their mechanism of size variation. Indeed, \cref{fig:newmetrics}(a) shows clearly that this mechanism is a mutual process among towns (at least two), as illustrated by two neighbouring towns, Thun and Thierachern, shown in \cref{fig:newmetrics}(a, b). Although the maximum size variation of Thun involves about twice as many nodes as for Thierachern, the two $|K_c(t)|$ curves are partially mirrored, which means many nodes lost by Thun are acquired by Thierachern. A third interesting point is related to the types of dynamics of $|K_c(t)|$. We observe towns exhibiting large size variations, reaching a sharp peak (maximum or minimum), and towns displaying  smaller jumps and plateaus (e.g. Bern). Also, the maximum size changes can occur at times $t \neq t_{MFE}$, even during the phase of flood withdrawal, as for the case of Thun and Thierachern. This is probably an effect of the spatio-temporal variability, implying that local peaks of road closures can occur earlier or later than the overall peak (whole river basin). In the case of Thun and Thierachern\textcolor{black}{,} this is an effect of the time lag of the lake Thun which needed to be filled first before being able to trigger inundations along the shorelines.

\cref{fig:newmetrics}(b) shows $L_c(t)$ for all the $67$ towns. As seen above, this measure can reach negative values, meaning that a town can gain nodes from neighboring ones, due to the rerouting mechanism. As in the previous plot, we group the $L_c(t)$ curves into three categories for better readability: towns undergoing losses, incorporating new nodes, or constant. Both prior to the flood onset and after the recovery phase, $L_c(t)$ is expected to be zero or close to zero, while in the interval between these two extremes, $L_c(t)$ is expected to take values $0<L_c(t)<1$ (for a town loss) or $-1<L_c(t)<0$ (for a town gain). Although this behaviour is observed for most of the towns in \cref{fig:newmetrics}(b), some of them (dotted lines) do not show the same tendency for $L_c(t)$ to return to zero during the flood's recovery phase, i.e. $\lim_{t \to t_{end}} L_c \neq 0$.  Indeed, the lake levels do not return to the initial level during the timeframe of the simulations but need a few days longer and some streets remain interrupted until the end of the simulation period. Also, \cref{fig:newmetrics}(b) confirms the finding of \cref{fig:newmetrics}(a) that the largest size variations of each town (the largest values of $L_c(t)$, both positive and negative) do not always occur at the maximum flood extension. For example, the largest positive value of $L_c(t)$, recorded among all the towns, occurs around $t=60$ hours and belongs to the town of Gelterfingen, around halfway between Bern and Thun. The largest negative value, instead, was recorded for the town of Thierachern a few hours after the maximum flood extension. This means that emergency services and assistance should be dynamically allocated over the flood timespan. In this specific example, Gelterfingen would have a priority over Thierachern, or Thun, which both peak later. We also note the order of magnitude of the variations of $L_c$ which for some cities can reach values as large as $60\%$. Such a large increase potentially represents a very large logistic burden during the flood, a crucial information that needs to be integrated in risk management. 

The quantity $L_c(t)$ is a quantity which changes over time, and we define the expansion/shrinking rate as $\zeta_c(t)=|L_c(t_{MFE})|$ (see Methods for detail). This quantity can be used for each town as an overall indicator of the size variation during the flood's growing phase, and gives an aggregated and more concise information about the impact of the flood on a given city. We show in \cref{fig:newmetrics}(c) the value of this indicator and we thus observe that some cities will indeed experience an important growth which is of utmost importance for preparation planning as they will constitute a set of resilient towns serving a large number of individuals. 
We also note that there are some small discrepancies between $L_c(t)$ and $\zeta_c$ which are due to the peaks (or in general to the largest values) of $|K_c|$ which can occur at  a different time than $t_{MFE}$. For example, the two towns which exhibit the largest values of $L_c(t)$, both positive and negative, i.e. Gelterfingen and Thierachern, do not score the largest values of $\zeta_c$ (see \cref{fig:newmetrics}(c)).

Rerouting entailed by a road disruption generally results in a longer path than the initial one (see \cref{fig:rerouting_mechanism}). In other words, \textcolor{black}{for a node $i$, the distance to its closest town $c'$ at time $t$, i.e. after rerouting, is longer than the distance to its closest town at $t_0$}: $d_{ic'}(t) \geq d_{ic}(t_0)$. A simple way to characterise the detour experienced by an individual living close to \textcolor{black}{a} city $c$ is then to compute the quantity $\eta_i(t) = \frac{d_{ic'}(t)}{d_{ic}(t_0)}-1$, in analogy to the \quotes{detour index} \cite{Barthelemy,Aldous}. The larger $\eta_i(t)$, and the larger the rerouting for connecting the node $i$ to its closest town's center $c'$ (which is different from the initially assigned town's center $c$). At each time $t$ of the disruptive event, the average rerouting from an initially assigned town center to a new closest one, can be captured by the average detour defined as
\begin{equation}
\overline{\eta}(t) = \frac{1}{m}\sum_{i=1} \bigg( \frac{d_{ic'}(t)}{d_{ic}(t_0)}-1\bigg) 
\label{averagedetour1}
\end{equation}
where $m$ is the number of nodes. This quantity\textcolor{black}{,} \cref{averagedetour1}\textcolor{black}{,} is a measure of the global cost or (in)efficiency of a network undergoing a disruptive event, since the rerouting would represent a time consuming and costly workaround to roads blockage. The global metrics $\overline{\eta}(t)$ and the loss averaged over all cities\textcolor{black}{,} $\overline{L}(t)=1/n\sum_{c=1}^nL_c(t)$, are shown in
\textcolor{black}{\cref{fig:newmetrics}(d, top) and \cref{fig:newmetrics}(d, bottom), respectively.}
Both these metrics are global indicators of the entire network performance, and, in contrast with $P_{\infty}$, they encode the local information coming from the entire network. Large values of $\overline{\eta}(t)$ and $\overline{L}(t)$ correspond to a large rerouting and a large loss of nodes for most towns. Ideally, the most efficient and robust network would therefore display a value $|\overline{\eta}(t) - \overline{\eta}(t_0)| = 0$, and $|\overline{L}(t) - \overline{L}(t_0)| = 0$, for all $t$. Despite sudden and abrupt variations of $\overline{\eta}(t)$ during the real-like flood, the network shows a significant higher efficiency than in a random flood scenario which causes large and gradual variations of $\overline{\eta}(t)$ over the entire flooding event. Surprisingly, the high efficiency of the network in the real-like flood appears simultaneously with a lower robustness (represented by $\overline{L}(t)$) than in the random flood which suggests that a low rerouting can be achieved through a larger redistribution of the town's Voronoi cells. However, the difference in $\overline{L}(t)$ between the two types of floods is not so pronounced as for $\overline{\eta}(t)$. A last remarkable difference between the two flood scenarios can be observed in the latest hours of the recovery phase: in the random flood, $\overline{L}(t)$ and $\overline{\eta}(t)$ tend to zero \textcolor{black}{(even though $\overline{\eta}(t)$ does not arrive to zero)}, while in a real-like flood, they reach a non-zero positive value (and a peak shifted in time)\textcolor{black}{. Basically, this is} due to the retention effect of lakes that causes flooding on roads later in time\textcolor{black}{, and which lasts for a few additional days up to return of the lakes' water levels to their initial ones}.

\section*{Discussion}
\label{Discussion}

The discussion on a realistic flood simulation presented here allowed us to show the inadequacy of percolation theory for evaluating the robustness of a transportation network, when subject to a realistic flooding event. 
\textcolor{black}{
The standard percolation process is characterised by a simultaneous and independent potential removal, usually using a probability $1-p$, of the entire set of edges. This means that all the edges undergo the removal process given by a probability $1-p$, and not just a subset. In this framework,
}
the size of the giant component, $P_{\infty}$ is a decreasing function of the growing fraction of removed nodes (or links) up to a certain threshold that depends on the specific network and where $P_{\infty}=0$. In a real-like flooding, 
\textcolor{black}{
instead,
}
the process of water propagation through the network can be thought as a localized attack both in time and space, 
\textcolor{black}{
where only a subset of the network’s edges can be affected. Therefore, the network cannot be completely disrupted if it is larger than the flooded area.
}
The realistic simulation of the flooding shows that, in general, we don’t observe $P_{\infty}=0$, even in an extreme event and at the peak of the flooding. 
\textcolor{black}{
Possible uncertainties in a flood propagation would not affect our observation, since a realistic mechanism of edges removal would not lead to $P_{\infty}=0$, if the considered network is large enough. Therefore,
}
the absence of an observed transition limits severely the relevance of percolation for studying this type of catastrophic event. In addition, we showed that the relevance of the giant component is debatable and we introduced metrics based on the idea that the crucial point for individuals during a disaster is the possibility to reach a town, even if it is not the closest one in the normal regime (see \textcolor{blue}{Supplementary Figure 1} for a representative visual comparison between the percolation approach and one of the proposed metrics).
\\
\\
The approach presented here answers questions relevant to flood risk management such as the spatio-temporal aspects of the flood’s impacts and where and when resources should be allocated during a flooding event. Our study displays the whole dynamics of the flooding and the sequence of important events, the location and time of the first relevant impact, how the road closures spread, which towns become less accessible, etc. In addition, our metrics about the dynamics of rerouting and rising/shrinking towns contains the relevant information for risk management decision makers, in order to allocate resources in space and time (e.g. hospitalization services, food supply, etc.). Already at the earliest moments of the flood, it could be beneficial to transfer resources to towns with a low potential of inundation and a large expansion/shrinking ratio $\zeta_c$.
\\
\\
\textcolor{black}{
Our study  showed that the standard percolation approach might have limits of applicability to real-world systems, such as in this case of flood-induced disruption of a single road network. Future studies might investigate whether such limits could occur in other real-world scenarios and potentially provide a more general theory of percolation or a more formal description of its limits of applicability. For example, still in transportation networks, what happens if we apply the percolation framework to disrupted multi-layer networks \cite{Kivela2014, Bianconi2018} which include road networks? Also, how can we address the issue related to the border size selection? Indeed, the size of the border determines the number of opportunities for deviating routes: a tiny border would eliminate many alternative routes, whereas a large border would enable a large number of possible alternatives. Therefore, future studies could investigated how far or how close the system can be delimited without losing potential routing alternatives. Also, the influence of the network topography on the flood’s impact could be further investigated in the future. Indeed, in our case, we studied the flood impacts on a road network situated in mountainous and hilly regions, while other authors considered quite flat areas, as 
Abdulla et al. \cite{Abdulla2019_conference_paper, Abdulla2020_conference_paper, Abdulla2020_SIS, Abdulla2021} with the Houston road network. Another direction for future works is related to a simple model for deriving \cref{eq:generalized_Verhulst_equation} and understanding its parameters. Such a simple mechanistic model might include basic principles of hydrology and hydrodynamics and, at least, should account for the difference in height among each pair of nodes, in order to provide the water flows directions. This simple model might also fulfil the criteria of universality, i.e. being applicable to a vast variety of scenarios and road networks. Indeed, so far, the 
Abdulla et al. \cite{Abdulla2020_SIS} and Fan et al. \cite{Fan2020} models of flood propagation (in a road network) seem to be case-dependent, and their parameters have to be estimated for any network and scenario. Another limitation that could be addressed in future studies is related to the use of realistic time-varying travel times. Indeed, travel times and the distribution of traffic congestion change during a flood and the corresponding network weights (represented by travel times) should be updated at each time step, in order to get more reliable results.}

\section*{Methods}

\subsection{Definition of town center on a transportation network}
\label{towncenter}

The definitions of the new proposed measurements (alternative to percolation) are built up around the concept of town center (or city center), which could be intuitively thought as the core area of a town, for social, business, and cultural activities. Despite the common idea of an extended area, the town center was here redefined as a node of a road network, as representative spot for the town center. For each Swiss municipality, two entities were employed for selecting the proxy node of a town center, i.e. the road network and polygons called (town or city) \emph{central zones} by the Swiss Federal Office \cite{SwissFederalOffice}. 
Since several Swiss towns included a multitude of those polygons, which were all named as central zones, the largest polygon was selected as \emph{town center}, often corresponding to the historical town center, while the other smaller polygons were neglected. The largest polygon was chosen by assuming it could include the largest quantity of services, useful for a population. Then, the roads overlapping and intersecting the largest polygon 
were selected, and the centroid of their endpoints was employed to infer the proxy node. The node representing the town center, 
was defined as the closest node to the roads endpoints' centroid, calculated from the roads which overlapped the largest polygon. Obviously, other definitions of town center's node are possible and can be used. For example, the small central zones and their overlapping roads could be also included in calculating the roads endpoints' centroid, but with the risk of obtaining a centroid far from the largest polygon.

\subsection{Definition of town and temporal evolution of a flood on a transportation network}
\label{town_and_flood_evolution} 

\begin{figure*}
\centering
\includegraphics[width=6.0cm]{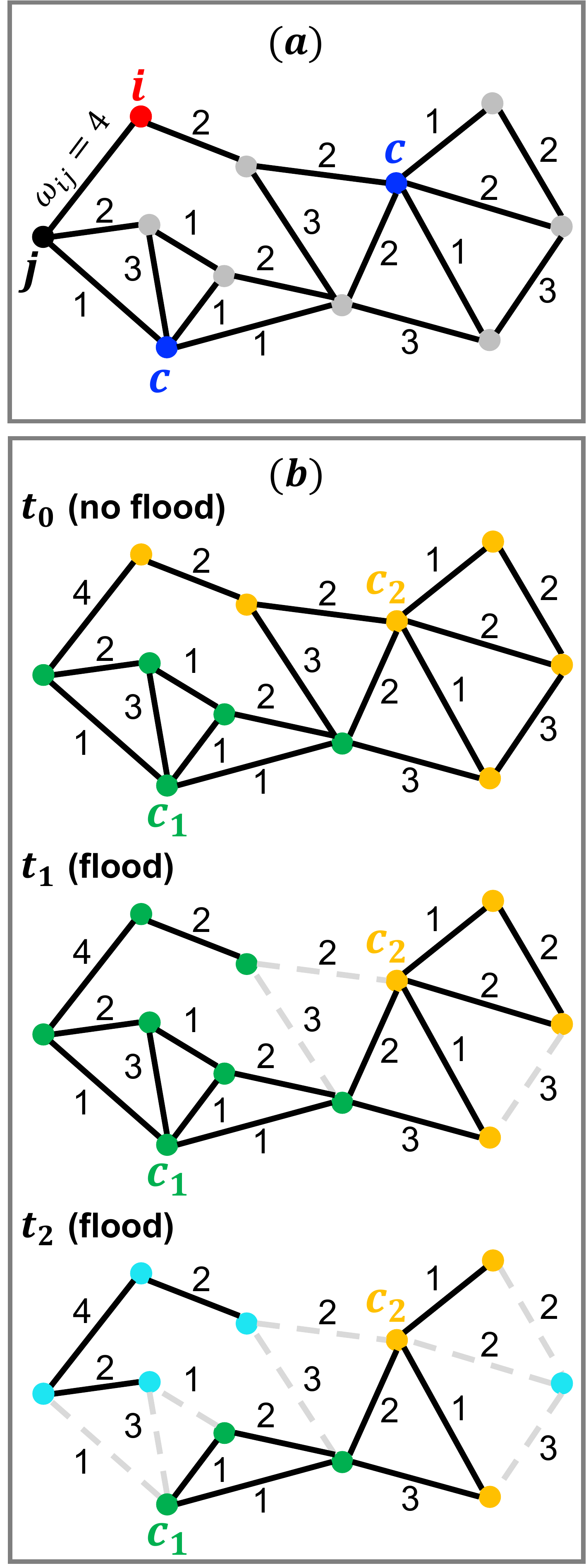}
\caption{Temporal evolution of towns during a flood event. \textbf{(a)}: Part of a weighted road network, illustrating a typical network's $i-$node and two $c-$nodes, which represent two town centers. Single-road lengths $d_{roads}$ and single-road travel times $\tau_{roads}$ were used as network weights.  \textbf{(b, top)}: Definition of towns, i.e. assignment of the road network’s nodes to the nearest town centers at $t_0$.  In \textbf{(b, middle)}, all nodes \quotes{tend to reach} the closest town centers, even if different from their initially assigned ones at time $t_0$. That means that the two nodes at the top left corner, which were yellow at $t_0$, become green at $t_1$, since now closer to the town center $c_1$. As consequence, the towns' sizes change over time.  In \textbf{(b, bottom)} further flood advancement wipes out other roads, isolating a number of nodes (as singlets or small clusters) from their closest town centers. In this example, at time $t_0$, the size of the two towns would be $|K_{1}(t_0)| = 5$, $|K_{2}(t_0)| = 6$. At time $t_1$, we would count $|K_{1}(t_1)| = 7$, $|K_{2}(t_1)| = 4$, while at time $t_2$, we would have $|K_{1}(t_2)| = 3$, $|K_{2}(t_2)| = 3$, and a total number of disconnected nodes equal to $N_\infty(t_2) = 5$.
\textcolor{black}{This figure was produced with PowerPoint \cite{PowerPoint2021}.}}
\label{fig:rerouting_mechanism}
\end{figure*}

\begin{figure*}
\centering
\includegraphics[width=15.6cm]{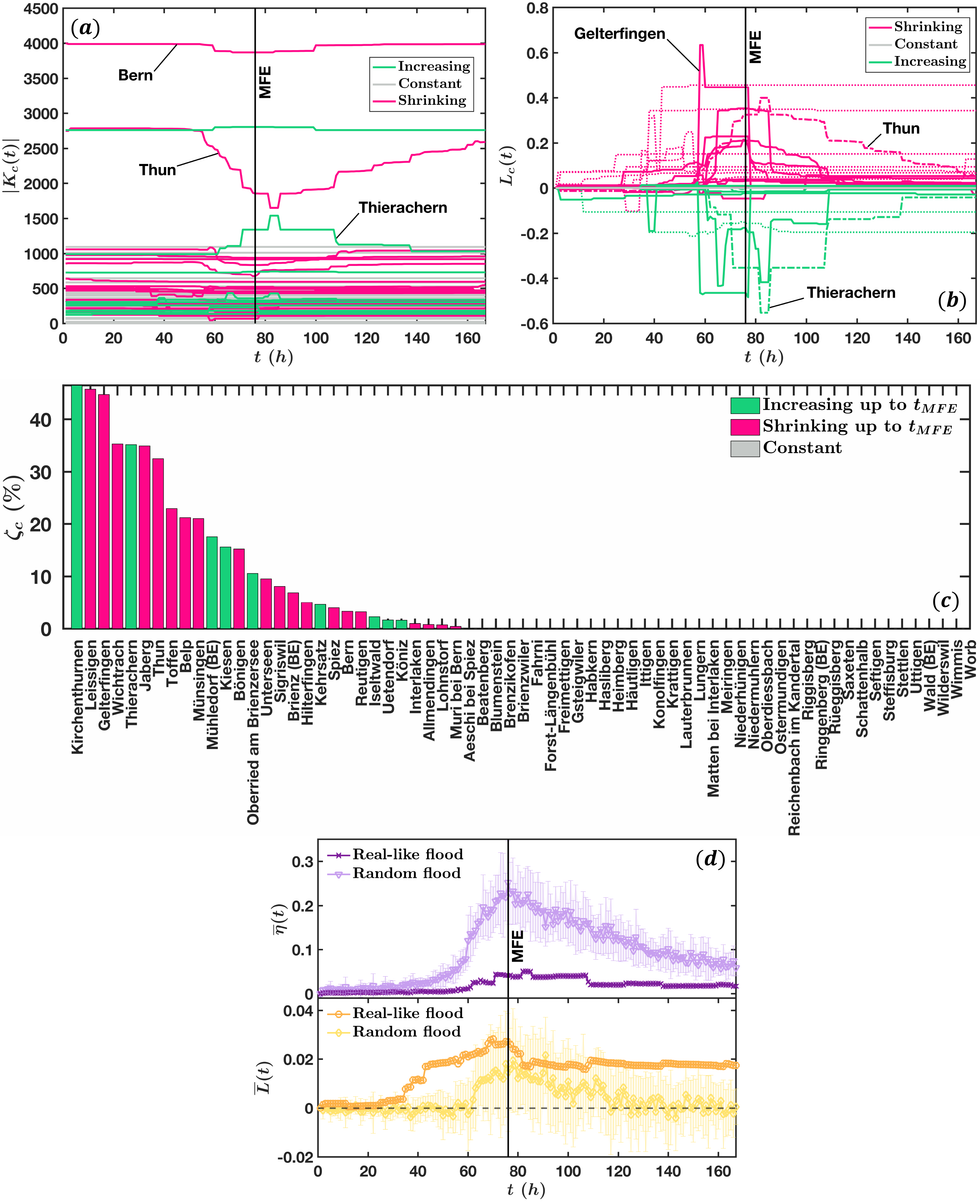}
\caption{Temporal evolution of \textbf{(a)} the \emph{town's size}, $|K_c(t)|$ (\cref{widetildeK_c}), \textbf{(b)} the \emph{town's loss}, $L_c(t)$ (\cref{loss}), \textbf{(c)} the town \emph{expansion/shrinking ratio} $\zeta_c$ (\cref{expansionratio}) and \textbf{(d)} the \emph{average detour} $\overline{\eta}(t)$ (\cref{averagedetour1}) together with the \emph{average town loss} $\overline{L}(t)$ (\cref{averageloss}). The local metrics $|K_c(t)|$, $L_c(t)$ and $\zeta_c$ capture the dynamics of each of the $67$ towns composing the network (with a perimeter of $5000$ meters). Due to the re-routing mechanism illustrated in \cref{fig:rerouting_mechanism}, the towns size increases, diminishes or remains constant. 
For a better visualisation we grouped the towns with \emph{overall size increase} in green, (i.e. if $|\max_{t}|K_c(t)| - |K_c(t_0)|| > |\min_{t}|K_c(t)| - |K_c(t_0)||$), the towns with \emph{overall size decrease} in fuchsia (i.e. if $|\max_{t}|K_c(t)| - |K_c(t_0)|| < |\min_{t}|K_c(t)| - |K_c(t_0)||$) and the towns without size variations in gray (i.e. if $|\max_{t}|K_c(t)| - |K_c(t_0)|| = |\min_{t}|K_c(t)| - |K_c(t_0)|| = 0$). The global metrics $\overline{\eta}(t)$ and $\overline{L}(t)$, representing the network's (in)efficiency and robustness, respectively, are here illustrated for both flood scenarios. For randomly generated floods, the average values of $\overline{\eta}(t)$ and $\overline{L}(t)$ were calculated over $20$ realizations.
\textcolor{black}{This figure was produced with Matlab \cite{MATLAB2019}.}}
\label{fig:newmetrics}
\end{figure*}

The town centers' nodes previously described in \cref{towncenter} (\nameref{towncenter}) represent an essential ingredient for the new metrics definitions. All those metrics are indeed based on the shortest distances $d_{ic}$ and on the quickest paths $\tau_{ic}$ among all the pairs of $i-$nodes of the road network and $c-$nodes of the town centers. At each time, we thus have matrices $d_{ic}$ and $\tau_{ic}$ where $i = 1,\dots,m$, $c = 1,\dots,n$. The quantities $m$ and $n$ represent the total numbers of road network's nodes and town center's nodes, respectively. 
The shortest distances and the quickest travel times were calculated with the Dijkstra algorithm \cite{Dijkstra}, by employing single-road lengths $d_{road}$ $[m]$ and single-road travel times $\tau_{road}$ $[s]$ as network weights (see \cref{fig:rerouting_mechanism}(a)). The single-road lengths were collected from the Swisstopo database \cite{Swisstopo}, while the single-road travel times were approximated by the following formula:
\begin{equation}
    \tau_{road} = \frac{d_{road}}{f \frac{v_{lim}}{3.6}}
    \label{tauroad}
\end{equation}
where, $f=0.8$ is the reduction speed factor which considers the relationships between speed limits and traffic flows \citep{Gao2019}, $v_{lim}$ is the speed limit $[km/h]$ from 
\textcolor{blue}{Supplementary Table 2},
and the number $3.6$ was used to convert $v_{lim}$ from $[km/h]$ into $[m/s]$.
Once the elements $d_{ic}$ and $\tau_{ic}$ are computed, the minimum value in each row of
\textcolor{black}{the corresponding matrices $D$ and $\Tau$}
is selected, i.e. the town center's node with minimum distance to the $i^{th}-$node. This corresponds to assign an $i^{th}-$node to the closest town center's node, at $t_0$. As a consequence, at time $t_0$, all nodes in the network are associated to their closest town centers' nodes (\cref{fig:rerouting_mechanism}(b, top)), creating $n$ sets of nodes
$K_c(t_0) = \{ i \mid \min\limits_{c'} (d_{i c'} (t_0)) = d_{ic} (t_0) \}$. These sets, $K_1,K_2,\dots,K_n$, correspond to Voronoi cells on the network and could be written also for travel times $\tau_{i c}$. We defined the generic $c-$th \emph{town} as:
\begin{equation}
    K_c \equiv K_c(t_0) \doteq \{ i \mid \min\limits_{c'} (d_{i c'} (t_0)) = d_{ic} (t_0) \}
    \label{K_c}
\end{equation}
such that $|\bigcup_{c=1}^{n} K_{c}| = m$. During the flood, the elements of $D$ and $\Tau$ could change due to the disruptions of the road network and we recompute them at each time step $t_0,t_1, t_2,\dots$, where $t_0$ is the time just before the flood start, with no disruptions (\cref{fig:rerouting_mechanism}(b, top)), and $t_1$ is the first time with disruptions (\cref{fig:rerouting_mechanism}(b, middle)). 
During the flood, some roads are disconnected and some nodes have to be rerouted, resulting in a reorganization of the Voronoi cells of each town.
We indicate the size of a town $c$ which can be variable by
\begin{equation}
K_c(t) \doteq \{ i \mid \min\limits_{c'} (d_{i c'} (t)) = d_{ic} (t) \}
    \label{widetildeK_c}
\end{equation}

\subsection{New metrics}
\label{newmetrics_methods}

\subsubsection*{Detour index}

Rerouting entailed by a road disruption generally results in a longer path than the initial one, 
i.e. $d_{ic'}(t) \geq d_{ic}(t_0)$, 
where $c' \neq c$ is the closest town's center at time $t$, to which the $i-$node would be reassigned. Then, one of the simplest ways to measure the deviation of a path's distance at time $t$ from its original value at $t_0$, would be straightforwardly the ratio between the two distances, or the \emph{detour}, $\eta_i(t) = \frac{d_{ic'}(t)}{d_{ic}(t_0)}-1$, in analogy to the \quotes{detour index} \cite{Barthelemy,Aldous}. Since $d_{ic'}(t) \geq d_{ic}(t_0)$, we would measure $\eta_i(t) \geq 0$ $\forall i$, with $\eta_i(t) = 0$ indicating the absence of rerouting to the closest neighbouring town. Therefore, the larger the $\eta_i(t)$, the larger the rerouting for connecting the $i-$node to the closest town's center $c'$ (which is different from the initially assigned town's center $c$). At each time $t$ of the disruptive event, the average rerouting from an initially assigned town center to a new closest one, can be captured by the \emph{average detour}, defined as in \cref{averagedetour1}.
We would consider \cref{averagedetour1} as a measure of global (in)efficiency of a network undergoing a disruptive event, since the rerouting would represent a time consuming and costly - i.e. inefficient - workaround to roads blockage. However, $\overline{\eta}(t)$ would differ from the well-established Latora \& Marchiori's \cite{Latora} or from the Vragović's \cite{Vragovic} global network efficiencies, which are basically averages of $\frac{1}{d_{ij}}$ and $\frac{d_{ij}^{Euclid.}}{d_{ij}}$, respectively, over all pairs of nodes. Indeed, the average in \cref{averagedetour1} would be only over those pairs of nodes which include the towns' center nodes, and not over all pairs of nodes.

\subsubsection*{Disconnected nodes}

The distances and their variations over time, as well as the relative travel times, can also be employed to quantify the \emph{total number of disconnected nodes} in the entire network:
\begin{align}
N_\infty(t) &= \sum\limits_{\substack{i=1}}^{m} b_{ic} 
\label{infinitedistances}
    \\
    \text{where} \quad b_{ic} &= \nonumber
\begin{cases}
    1,& \text{if } d_{i c} (t) = \infty \quad (\text{or} \; \tau_{i c} (t) = \infty)\\
    0,              & \text{otherwise}
\end{cases}
\end{align}

\subsubsection*{Town loss}

Our understanding of the flood’s effects can be further enriched by metrics related to the size’s variations of each town, $|K_c(t)|$. The size reduction of a town during a flood is intuitively expected and can be expressed, analogously to $\eta_i(t)$ for the distances, through a ratio between the size of the damaged (and diminished) town at time $t$, and the original size at time $t_0$:
\begin{equation}
L_c(t) = 1 - \frac{|K_c(t)|}{|K_c|} 
\label{loss}
\end{equation}
For example, we can have a reduction of the town size from an initial $|K_c|=2500$ nodes to $|K_c(t)|=2000$ nodes at time $t$, or rather a size loss equal to $L_c(t)=0.2$, i.e. a $20\%$ less than the initial size at time $t_0$. In parallel to $\eta_i(t)$, we can also define a global measure for $L_c(t)$, the \emph{average town loss}, simply by averaging $L_c(t)$ over all towns:
\begin{equation}
\overline{L_c}(t) = \frac{1}{n}\sum_{c=1} \bigg(
1-\frac{|K_c(t)|}{|K_c|} \bigg) 
\label{averageloss}
\end{equation}

\subsubsection*{Expansion/shrinking ratio}

While \cref{loss} provides information at each time $t$ during the entire flood event, the size variations of every town can be even characterized by a single representative value, which relates the town size at the time of the maximum flood extension\textcolor{black}{,} $t_{MFE}$\textcolor{black}{,} to the town size at $t_0$. We call it the \emph{expansion/shrinking rate}:

\begin{equation}
\zeta_c  = \frac{\displaystyle\left\lvert|K_c(t_{MFE})| - |K_c|\right\rvert}{|K_c|}
\label{expansionratio}
\end{equation}
Although we would expect the maximum network's disruption at $t_{MFE}$, where the flood reaches the maximum extension, it might be that some towns could reach their maximum disruption at another time $t \neq t_{MFE}$. Therefore, $\max\limits_{t} \big(|K_c(t)|\big)$ might be different from $|K_c(t_{MFE})|$, and \cref{loss} and \cref{expansionratio} could eventually indicate different size variations for the same town.

\subsection{Flood failure}
\label{flood_failure}

\textcolor{black}{A realistic and extreme flood scenario \cite{Zischg} based on physically plausible rainfall scenarios and a high spatio-temporal resolution was used to generate the road network disruptions. The spatio-temporal patterns of rainfall of a three-day probable maximum precipitation event were generated with a Monte Carlo procedure that considers only physically plausible patterns. The hydrological model simulated the outflows from the subcatchments. A 1D hydrodynamic flood model simulated the water fluxes from the tributaries through the main river network and the lakes. The 1D hydrodynamic model provided the boundary conditions for the 2D flood simulation models that had been set up for each floodplain. The 2D hydrodynamic model had a spatial resolution of 50 m and computed the flood depths for each timestep (hourly)  and thus provided the necessary information for the impact model. The impact model simulated the water depths at the road edges. The model considers the location of the bridges in the third dimension, i.e., road bridges located above the simulated water surface elevation are considered as not affected by the flood. In addition, the model has been validated with historic events.}

\bibliography{ms_SciRep_Loreti_V2_cleaned_arXiv}

\section*{Acknowledgements}

S.L. thanks Alexander Krok, Maria Arduca and Riccardo Turin for useful discussions. S.L. thanks the Mobiliar Lab for Natural Risks for funding and supporting this research project.

\section*{Author contributions statement}

Author contributions: S.L., A.Z., M.K. and M.B. designed research; S.L. performed research; S.L., E.S.-G., A.Z., M.K. and M.B. analyzed data; S.L. and M.B. wrote the paper.

\section*{Additional information}

\textbf{Competing interests:} The authors declare no competing interest.



\end{document}


\maketitle
%
%


\vspace{-0.5cm}
\noindent 
\textbf{Date:}
31 January 2022.\\
\textbf{Disclaimer:}
This research article has been published the 28th of January 2022 on \href{https://www.nature.com/articles/s41598-022-04927-3 }{Scientific Reports}.\\
\textbf{DOI:} https://doi.org/10.1038/s41598-022-04927-3.\\
\textbf{Cite this article:}
Download citation from the \href{https://www.nature.com/articles/s41598-022-04927-3#citeas}{Scientific Reports} citation tool.
\vspace{0.5cm}

\section*{Supplementary Text}
\subsection*{Generalized logistic function}


We consider the generalised Verhulst's equation \cite{Verhulst},
\begin{align}
\frac{dF(t)}{dt} &= \beta_{1} \biggl(F-C_1\biggr) \biggl(1- \biggl(\frac{F-C_1}{D_{1}} \biggr)^{\gamma_1} \biggr)
\intertext{where $C_1$ is the lower asymptote, $D_1$ is the upper asymptote, $\gamma_1$ is an exponent which allows to vary the shape of the (solution) sigmoidal curve and $\beta_1$ is the \emph{intrinsic growth rate} \cite{Tsoularis} indicating the curve steepness. We then derive the 
corresponding generalised logistic function as follows:}
\beta_{1} \int_{t_0}^t dt &= \int \frac{dF}{\biggl(F-C_1\biggr)\biggl(1- \biggl(\frac{F-C_1}{D_{1}} \biggr)^{\gamma_1} \biggr)} \nonumber
\\
\beta_{1} (t-t_0) &= ln\biggl|F-C_1\biggr| - \frac{ln\biggl|1-\biggl(\frac{F-C_1}{D_{1}}\biggr)^{\gamma_1}\biggr|}{{\gamma_1}} + \text{const.} \nonumber
\\
{\gamma_1} \beta_{1} (t-t_0) &= ln\biggl|F-C_1\biggr|^{\gamma_1} - ln\biggl|1-\biggl(\frac{F-C_1}{D_{1}}\biggr)^{\gamma_1}\biggr| + \text{const.} \nonumber
\\
- {\gamma_1} \beta_{1} (t-t_0) &= - ln\biggl|F-C_1\biggr|^{\gamma_1} + ln\biggl|1-\biggl(\frac{F-C_1} {D_{1}}\biggr)^{\gamma_1}\biggr| + \text{const.} \nonumber
\\
- \text{const} - {\gamma_1} \beta_{1} (t-t_0) &= ln\left| \frac{1-\biggl(\frac{F-C_1} {D_{1}}\biggr)^{\gamma_1}}{(F-C_1)^{\gamma_1}} \right| \nonumber
\\ \intertext{we use $W = \pm e^{- \text{const}}$,}
W e^{- {\gamma_1} \beta_{1} (t-t_0)} &= \frac{1-\biggl(\frac{F-C_1} {D_{1}}\biggr)^{\gamma_1}}{(F-C_1)^{\gamma_1}} \nonumber
\\
W e^{- {\gamma_1} \beta_{1} (t-t_0)} &= \biggl( 1-\biggl(\frac{F-C_1} {D_{1}}\biggr)^{\gamma_1} \biggr) \frac{1}{(F-C_1)^{\gamma_1}} \nonumber
\\
W e^{- {\gamma_1} \beta_{1} (t-t_0)} &= \frac{1}{(F-C_1)^{\gamma_1}} - \frac{1}{D_{1}^{\gamma_1}} \nonumber
\\
\frac{1}{D_{1}^{\gamma_1}} + W e^{- {\gamma_1} \beta_{1} (t-t_0)} &= \frac{1}{(F-C_1)^{\gamma_1}} \nonumber
\\
\frac{1 + D_{1}^{\gamma_1} W e^{- {\gamma_1} \beta_{1} (t-t_0)}}{D_{1}^{\gamma_1}} &= \frac{1}{(F-C_1)^{\gamma_1}} \nonumber
\\ \intertext{we use $A_{1} = D_{1}^{\gamma_1} W$,} 
(F-C_1)^{\gamma_1} &= \frac{D_{1}^{\gamma_1}}{1 + A_{1} e^{- {\gamma_1} \beta_{1} (t-t_0)}} \nonumber
\\
F-C_1 &= \frac{D_{1}}{\biggl( 1 + A_{1} e^{- {\gamma_1} \beta_{1} (t-t_0)}\biggr)^{1/{\gamma_1}}} \nonumber
%
\\ \intertext{Finally, we obtain the generalised logistic function:} 
F(t) &= \frac{D_{1}}{\biggl(1+A_{1}e^{-\beta_{1}{\gamma_1}(t-t_0)}\biggr)^{1/{\gamma_1}}} + C_{1}
\label{extended_logistic_function}
\end{align}
With lower asymptote equal to $C_1=0$, \cref{extended_logistic_function} would be solution of the Richards differential equation \cite{Nelder,Richards,Wang2012,Turner}.








%
%
\section*{Supplementary Figures}
%
%
\begin{figure}[H]
\centering
\includegraphics[width=12cm]{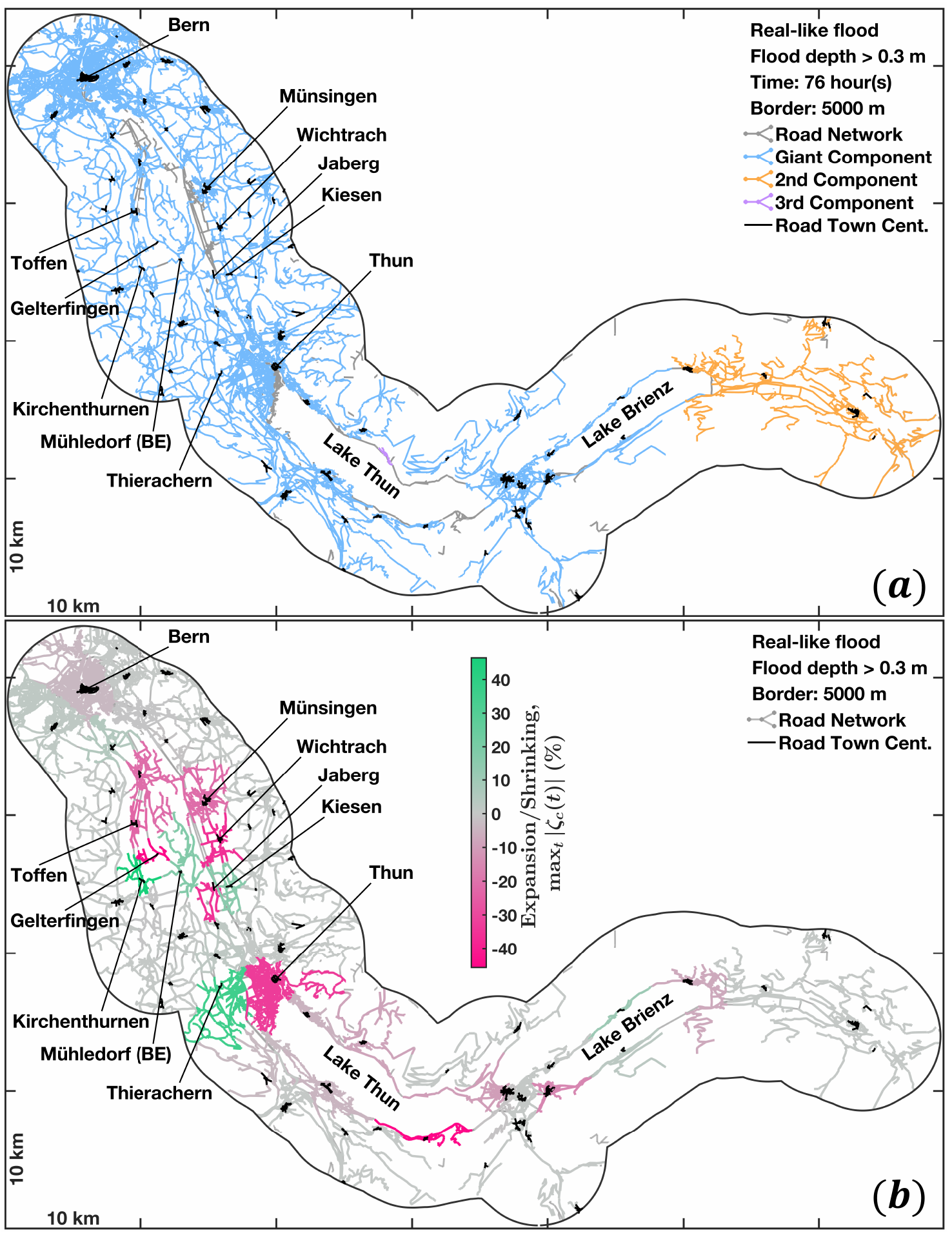}

\caption{A representative visual comparison between the percolation approach and our proposed framework and metrics.
\textbf{(a)} Snapshot of the road network and of the three largest connected components during the maximum extension of a real-like flood \citep{Zischg}, at $t_{MFE}=76$ hours. The percolation approach shows that the mobility of people occurs within each of the clusters, but without the possibility of crossing from one cluster to another since they are disconnected. However, we do not have any information about the ``internal'' situation of each cluster. For example, it looks like that (apparently) vehicles can freely circulate without any obstacle or speed braking within the giant component.
\textbf{(b)} Illustration of the road network, with highlight on the maximum value of $|\zeta_c|$ over the entire flooding period (for each town). With our approach, differently by percolation, we are able to infer the internal dynamics of each cluster, identifying the pivotal towns where large variations occur, both in space and in time. In particular, \textbf{(b)} shows the ``re-routing mechanism'' which corresponds to the mutual exchange of nodes among two or more adjacent towns (or groups of towns). We can clearly observe this mechanism between Thun and Thierachern, where Thierachern acquires some of the nodes lost by Thun, and between Kirchenthurnen and Gelterfingen, where Kirchenthurnen acquires some of the nodes lost by Gelterfingen. If we extend further our vision, we could observe the mechanism of mutual exchange of nodes among other adjacent towns, i.e. between the two groups of (i) Kirchenthurnen-Mühledorf and (ii) Gelterfingen-Wichtrach-Jaberg. Exception is made for the town of Münsingen, which just loses many nodes but there are not adjacent towns which acquire its lost nodes. 
\textcolor{black}{This figure was produced with Matlab \cite{MATLAB2019}.}
}
\label{fig:map_giantcomponent_76_hour}
\end{figure}

%
%
%
\section*{Supplementary Tables}

%
%
\begin{table}[H]
\centering
\begin{tabular}[t]{ c c c c c c c c c c}
\toprule
&$R^2$&$A$&$B$&$C$&$D$&$E$&$t_p$&$\beta$&$\gamma$\\
\midrule
$\frac{D_1}{\big(B_1 + A_1 e^{-E_1 (t-t_p)}\big)^{1/\gamma_1} } + C_1$ 
& $0.9959$ & $16.60$ & $0.97$ & $50.00$ & $834.70$ & $0.2295$ & $52.2$  & $0.145$ & $1.581$\\
$B_2 + A_2 e^{\beta_2 t}$ & $0.9975$ & $3.73$ & $35.89$ & & & & &  $0.0698$ \\
$A_3 e^{-\gamma_3 t}$ & $0.9981$ & $2334$ & & & & & & &  $0.0134$ \\

$B_4 + A_4 e^{\beta_4 t}$ & $0.9800$ & $64.94$ & $416.30$ & & & & &  $0.0434$ \\

$B_5 + A_5 e^{-\gamma_5 t}$ & $0.9911$ & $3836$ & $800.60$ & & & & & &  $0.0158$ \\
\bottomrule
\end{tabular}
  \caption{Best fit values.}
  \label{Tab:table_bestfit}
\end{table}

%
\begin{table}[H]
\centering
\begin{tabular}[t]{l>{\raggedright}p{0.35\linewidth}>{\raggedright\arraybackslash}p{0.3\linewidth}}
\toprule
&Interurban Speed Limit $[km/h]$&Urban Speed Limit $[km/h]$\\
\midrule
1 m road&-&50 \\
2 m road&80&50 \\
3 m road&80&50 \\
4 m road&80&50 \\
6 m road&80&50 \\
8 m road&80&50 \\
10 m road&80&50 \\
entrance&50&- \\
exit&50&- \\
highway&100&- \\
motorways&120&- \\
service area&20&- \\
service areas connection&20&- \\
service area entrance&40&- \\
square&-&20 \\
\bottomrule
\end{tabular}
  \caption{Speed limits adopted in this study for different road types and widths \cite{Finocchio}.}
  \label{Tab:speedlimits}
\end{table}%

\clearpage

\bibliography{supp_SciRep_Loreti_V2_SupplementaryInformation}